\documentclass[aps,prl,reprint,twocolumn,superscriptaddress]{revtex4-2}

\usepackage{graphicx,psfrag,amsmath,amssymb,amsfonts,bbm,latexsym,color,dcolumn,bm,mathbbol}
\usepackage[framemethod=tikz]{mdframed}
\usepackage[colorlinks=true,allcolors=blue]{hyperref}
\usepackage{soul} 
\usepackage{cancel}

\usepackage{array}
\usepackage{booktabs}
\usepackage{footnote}
\usepackage{threeparttable}

\usepackage{comment}

\usepackage{natbib}
\usepackage{amsmath,amssymb}             
\usepackage{color}

\usepackage{enumerate}

\definecolor{szlcolor}{rgb}{0.8, 0, 0.6}

\definecolor{bluecolor}{rgb}{0, 0, 1.0}

\definecolor{redcolor}{rgb}{1, 0, 0}

\usepackage[normalem]{ulem}
\definecolor{jfcolor}{rgb}{0.1, 0.0, 0.9}

\usepackage[normalem]{ulem}
\definecolor{jfcolor2}{rgb}{0.8, 0.0, 0.9}

\begin{document}

\title{Solid-to-solid transition in dense assemblies of elongated cells}

\author{Shao-Zhen Lin}
\email{linshaozhen@mail.sysu.edu.cn}
\affiliation{Guangdong Provincial Key Laboratory of Magnetoelectric Physics and Devices, School of Physics, Sun Yat-sen University, Guangzhou 510275, China}
\affiliation{Interdisciplinary Research Center for Physical Mechanics in Complex Systems and Its Engineering Applications, School of Physics, Sun Yat-sen University, Guangzhou 510275, China}
\author{Jean-François Rupprecht}
\email{jean-francois.rupprecht@univ-amu.fr}
\affiliation{Aix Marseille Univ, CNRS and Turing Centre for Living Systems, CPT (UMR 7332), Marseille, 13009 (France).}
\affiliation{Aix Marseille Univ, Université de Toulon, CNRS, LAI (UMR 7333), Turing Centre for Living Systems, Marseille, France}

\begin{abstract}
Cell shapes in confluent tissues range from nearly isotropic epithelial morphologies to highly elongated endothelial ones. 
In standard vertex models, tissue rigidity is controlled by a target shape index; increasing this index drives cell elongation and ultimate tissue fluidization. 
Here, we consider the case where cell elongation emerges autonomously by assigning an intrinsic, passive elastic preference for anisotropic shape.
This distinction reverses the usual expectation: cell elongation does not fluidize the tissue, but drives a solid-to-solid transition from an ordered isotropic solid to a disordered anisotropic solid, with finite yield stress and shear rigidity on either side of the transition. 
These results decouple cell shape from tissue rheology and caution against inferring fluid-like mechanics from elongated cell morphologies alone.
\end{abstract}

\date{\today}

\maketitle

\textit{Introduction.} -- Embryo folding, wound closure, and tumor invasion all depend on tissue mechanics, at the interface of cell biology and soft condensed matter physics \cite{Massey2024,Hayward2021,Goodwin2021,Trubuil2021,Autorino2026,Rustarazo-Calvo2026NP}.
Vertex models provide a minimal description of confluent cell sheets and capture jamming, fluidization, and collective rearrangements \cite{Staple2010,Hannezo2014,Bi2015,Merkel2019,Barton2017,Lin_PNAS_2017,Lin2018,Rozman2025,Staddon2025,Kim2024,Chen2022,Yu2025,Claussen2026}. In these models, tissue rigidity is controlled by the target shape index, the dimensionless ratio between a cell’s preferred perimeter and the square root of its preferred area. Increasing this index lowers the energy barriers to cell neighbor exchanges, which eventually leads to a solid-to-fluid transition \cite{Staple2010,Bi2015,Merkel2019}.

\begin{figure}[t!]
\centering
\includegraphics[width=8.0cm]{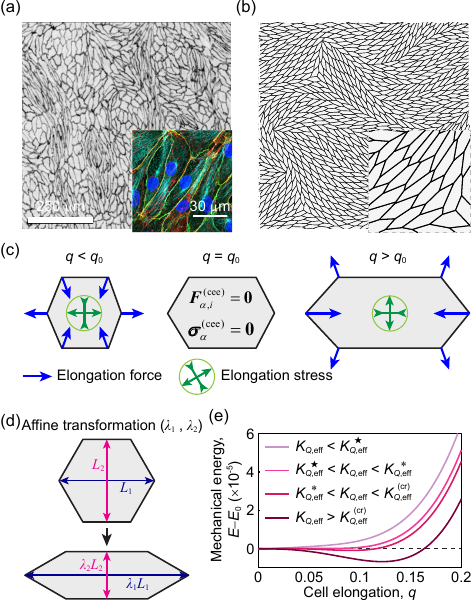}
\caption{\label{fig:model} 
Vertex model with cell elongation elasticity (CEE). 
(a) HUVEC monolayer under vascular endothelial growth factor. Adopted from Refs. \cite{Cao2017,Barrasa-Ramos2025}. 
(b) Vertex model simulation with CEE. 
(c) Sketch of the CEE, showing the cell elongation forces and stresses for different cell elongations $q$. 
(d) Sketch of the affine transformation $( \lambda_1 , \lambda_2 )$ applied to a hexagonal cell. 
(e) Cell mechanical energy $E$ versus cell elongation $q$ for different $K_{Q,\rm eff}$. 
The cell undergoes an affine transformation $(\lambda_1 , \lambda_2)$ with $\lambda_1 \lambda_2 = 1$; $E_0 = E (q = 0)$.
}
\end{figure}

Yet biological cell elongation need not arise from the isotropic area--perimeter competition that controls the standard rigidity transition \cite{Rauzi2008,Bi2015}.
Many cell types possess a preferred elongated morphology, set by cytoskeletal organization, polarity, adhesion, or environmental cues, as observed for instance in C2C12 myoblasts \cite{Duclos_NP_2017,Huo2025}, neural progenitor cells \cite{Kawaguchi_Nature_2017}, and human umbilical vein endothelial cells \cite{Cao2017,Barrasa-Ramos2025} [Fig.~\ref{fig:model}(a)], among other systems \cite{Duclos_NP_2018,Blanch-Mercader_PRL_2018,Tlili_PNAS_2019,Makhija2024}.
Here, this autonomous elongation is isolated through cell elongation elasticity (CEE), an energy cost for departing from a preferred anisotropic shape.

Our question is whether autonomous elongation fluidizes a tissue up to a solid-to-fluid transition, as suggested by shape-index arguments, or instead provides a distinct route to rigidity.

Our answer is that CEE drives a solid--solid transition in the vertex model of confluent cell sheets, from an ordered, isotropic solid state to a disordered, anisotropic solid state marked by changes in cell shape, tissue order parameter, shear modulus, and yield stress. 
This transition differs from previously reported jamming or fluidization transitions, establishing CEE as an independent control parameter for tissue mechanics. Together, these results extend vertex models to include CEE and supply a microscopic mechanism for the solid--solid transition in biological tissues, linking how cells set their own shape to how a tissue resists deformation during morphogenesis and beyond. 

\textit{Model.} -- A vertex model casts an epithelial monolayer as a polygonal tiling whose degrees of freedom are the vertex positions $\bm{r}_i$ (Fig. \ref{fig:model}(b)). 
Their overdamped dynamics balance friction against cell-shape relaxation forces derived from the tissue's mechanical energy: 
$E_{\rm svm} = \sum_{\alpha = 1}^{N_c} [ K_A \left( A_{\alpha} - A_0 \right)^2 / 2 + K_P \left( P_{\alpha} - P_0 \right)^2 / 2 ]$, which sums over cells $\alpha$ with area $A_{\alpha}$ and perimeter $P_{\alpha}$. 
This energy penalizes shape deviations from a reference state through the target area $A_0$ and perimeter $P_0$, yet a single ratio -- the target shape index $p_0 = P_0 / \sqrt{A_0}$ -- controls the tissue rheology: increasing $p_0$ fluidizes the tissue, rigidity vanishing above $p_0^\ast=2\sqrt{2\sqrt{3}}\simeq3.72$ for a hexagonal tiling and $p_0^\ast\simeq3.81$--$3.94$ in disordered Voronoi tissues \cite{Staple2010,Bi2015,Park2015,Merkel2019}. 

To model cells with an intrinsic tendency to elongate, we supplement the standard-vertex-model energy with a liquid-crystal-inspired elastic term that penalizes deviations from a preferred anisotropic cell shape $q_0$ \cite{Marchetti2013,Giomi2015,Doostmohammadi2018,Julicher2018},
\begin{equation}
E_{\rm cee} = \sum\limits_{\alpha = 1}^{N_c}{\frac{1}{2}{{K}_{Q}}{{\left[ \text{tr}\left( \bm{Q}_{\alpha}^{2} \right)-\frac{1}{2}q_{0}^{2} \right]}^{2}}} . 
\label{eq:ElongationEnergy}
\end{equation}
Here, $K_Q$ is an elongation stiffness; $\bm{Q}_{\alpha}$ a second-order tensor describing cell-shape anisotropy based on cell edges: $\bm{Q}_{\alpha} = ({1}/{P_{\alpha}}) \sum_{k \in cell \ \alpha} \ell_k \bm{t}_k \otimes \bm{t}_k - \bm{I}/2$, where $\ell_k = | \bm{r}_{k+1} - \bm{r}_k |$ and $\bm{t}_k = (\bm{r}_{k+1} - \bm{r}_k) / \ell_k$ are the length and direction of the $k$-th edge of the $\alpha$-th cell; $\bm{I}$ is the second-order unit tensor \cite{Sonam2023,Lin2023}. The resulting elongation forces $\bm{F}_{i}^{(\text{cee})} = -\partial E_{{\rm cee}} / \partial \bm{r}_i$ yield zero net force and torque (Fig. \ref{fig:model}(c)). 
Using Batchelor's formula \cite{Batchelor_JFM_1970,Lin2022}, we obtain the elongation stress (SM \cite{SM}, Sec. I), $\bm{\sigma }_{\alpha}^{( \text{cee} )} = [ {{{K}_{Q}}( q_{\alpha}^{2}-q_{0}^{2} )}/({4{{A}_{\alpha}}}) ] [ q_{\alpha}^{2}\bm{I} + 2( 2-q_{\alpha}^{2} ){{\bm{Q}}_{\alpha}} -4{{\bm{Q}}_{\alpha}}:{{\bm{T}}_{\alpha}} ]$, where $q_{\alpha} = \sqrt{2\text{tr}(\bm{Q}_{\alpha}^2)}$ quantifies cell elongation and ${{\bm{T}}_{\alpha}} = ({1}/{{{P}_{\alpha}}})\sum_{k \in cell\ \alpha}{{{\ell}_{k}}{{\bm{t}}_{k}}\otimes {{\bm{t}}_{k}}\otimes {{\bm{t}}_{k}}\otimes {{\bm{t}}_{k}}}$. 
Since $\text{tr} [ \bm{\sigma }_{\alpha}^{(\text{cee})} ] = 0$, the elongation forces ($\bm{F}_{i}^{(\text{cee})}$) and stresses ($\bm{\sigma }_{\alpha}^{( \text{cee})}$) do not contribute to cell area changes. 

We relax the tissue to its minimal-energy state through the frictional dynamics $\gamma\mathrm{d}\bm{r}_i/\mathrm{d}t = - \partial E / \partial \bm{r}_i$ with $\gamma$ the friction coefficient and $E = E_{\rm svm} + E_{\rm cee}$ the total mechanical energy and T1 transitions for cell--cell interfaces shorter than a length threshold $\ell_{\rm T1}$ \cite{Fletcher_BJ_2014,Lin2023} (SM \cite{SM}, Sec. III). 
We initialize the system as $N_c = 10^3$ cells on a regular hexagonal lattice in a square domain with periodic boundary conditions. 
We rescale the parameters by the length scale $\sqrt{A_0}$ and the stress scale $K_A A_0$, and hereafter, refer all parameters to the rescaled form, e.g., $P_0 \rightarrow P_0 / \sqrt{A_0}$. 
We set the rescaled parameters as in Table S1 \cite{SM}: $K_P = 0.02$, $P_0 = 1$ ($< P_0^{\ast} \approx 3.72$), $K_Q = 1$, $q_0 = 1$, and $\ell_{\rm T1} = 0.01$ if not otherwise stated. 

\begin{figure}[t!]
\centering
\includegraphics[width=8.6cm]{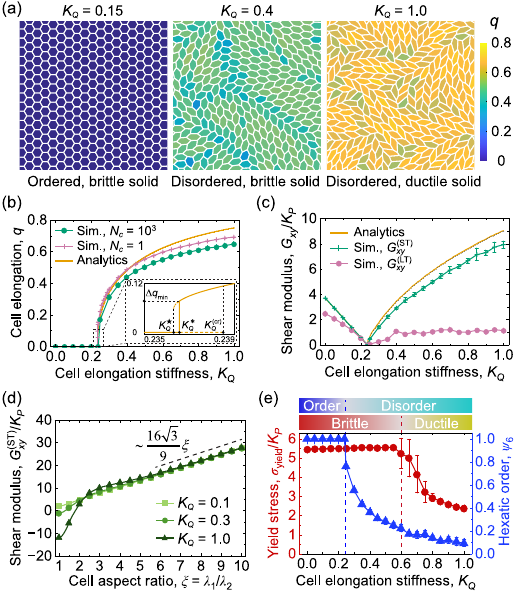}
\caption{\label{fig:S2S_transition} 
Autonomous cell elongation drives a solid-to-solid transition in a cell sheet for $P_0 < P_0^{\ast}$ with $P_0 = 1$. 
(a) Typical tissue morphologies at different $K_{Q}$. 
(b) Average cell elongation $q$ versus cell elongation stiffness $K_{Q}$. 
Symbols: numerical simulations. 
Line: theoretical prediction. 
Inset: $q_{\min}$ versus $K_Q$, where the solid line represents $q_{\min}$ and dashed lines refer to other local minima.
(c) Tissue shear modulus $G_{xy}$ versus cell elongation stiffness $K_{Q}$. 
Symbols: the short-time shear modulus $G_{xy}^{\rm (ST)}$ and the long-time shear modulus $G_{xy}^{\rm (LT)}$. 
Line: theoretical prediction of $G_{xy}^{\rm (ST)}$. 
(d) Predicted short-time shear modulus $G_{xy}^{\rm (ST)}$ versus cell aspect ratio $\xi = \lambda_1 / \lambda_2$ at different $K_Q$. 
(e) Tissue yield stress $\sigma_{\rm yield}$ and hexatic order parameter $\psi_6$ versus cell elongation stiffness $K_{Q}$. 
Parameters in Table S1 \cite{SM}. 
}
\end{figure}

\textit{Results.} -- At the single-cell level, CEE destabilizes the isotropic shape above a critical elongation stiffness, at which the cell shape undergoes a sharp jump. 
The mechanical energy $E$ of a single hexagonal cell reads
\begin{align}
E(q) \simeq \alpha_0 + \alpha_2 q^2 + \alpha_3 q^3 + \alpha_4 q^4,  \label{eq:E_approximation}
\end{align}
for small shape perturbations $q$ (SM \cite{SM}, Sec. II), with
$\alpha_2 = (K_{Q,\text{eff}}^{(\text{cr})}-{{K}_{Q,\text{eff}}}) / 4$, $\alpha_3 = - K_{Q,\text{eff}}^{(\text{cr})} / 18$, and $\alpha_4 = 2K_{Q,\text{eff}}^{(\text{cr})}/9 + {{K}_{P}}R_h{{P}_{0}}/3 + {{K}_{Q}}/8$ with $K_{Q, \rm eff} = K_Q q_0^2$ an effective elongation stiffness and ${R}_h$ the radius of a hexagonal cell at rest, satisfying $9{R}_{h}^{3} + ( 24{{{{K}}}_{P}}-2\sqrt{3} ){{{R}}_{h}}-4{{{K}}_{P}}{{{P}}_{0}} = 0$. 
$K_{Q, \rm eff}^{\rm (cr)}$ is a critical $K_{Q,\rm eff}$: 
\begin{equation}
{K}_{Q,\text{eff}}^{(\text{cr})} = 8{{{K}}_{P}}{{{R}}_{h}}\big( 6{{{{R}}}_{h}}-{{{{P}}}_{0}} \big) . \label{eq:Critical_KQ_eff}
\end{equation}
The sign of $\alpha_2$ sets the stability of an isotropic cell: when $K_{Q,\rm eff}$ exceeds $K_{Q,\rm eff}^{\rm (cr)}$, $\alpha_2$ turns negative and the cell elongates spontaneously. Yet, a cell shape transition occurs before ${K}_{Q,\text{eff}}^{(\text{cr})}$, as the cell elongation
$q_{\min}$ that minimizes $E(q)$, Eq. \eqref{eq:E_approximation}, undergoes a first-order-like transition at $K_{Q,\rm eff}^{\ast}$ ($<{K}_{Q,\text{eff}}^{(\text{cr})}$). 
Indeed, $q_{\min} = 0$ is the only minimum of $E(q)$ when $K_{Q,\rm eff} < K_{Q,\rm eff}^{\star}$; when $K_{Q,\rm eff}^{\star} < K_{Q,\rm eff} < K_{Q,\rm eff}^{\ast}$, two local minima exist, yet $q = 0$ remains the global minimum; when $K_{Q,\rm eff}^{\ast} < K_{Q,\rm eff} < K_{Q,\rm eff}^{\rm (cr)}$, two local minima exist but the global minimum switches to $q_{\rm e} = {3K_{Q,\text{eff}}^{( \text{cr} )}+3\sqrt{2( {{K}_{Q,\text{eff}}}-K_{Q,\text{eff}}^{(\text{cr})} )\nu+{{( K_{Q,\text{eff}}^{(\text{cr})} )}^{2}}}}/{\nu} > 0$, with $\nu = 32K_{Q,\text{eff}}^{(\text{cr})} + 48{{K}_{P}}R_h{{P}_{0}} + 18{{K}_{Q}}$; when $K_{Q,\rm eff} > K_{Q,\rm eff}^{\rm (cr)}$, $q_{\rm e} > 0$ is the only minimum; see Fig. \ref{fig:model}(e) and Fig. S3 for these different cases. 
These critical values of $K_{Q,\rm eff}$ read: $K_{Q,\rm eff}^{\star} < K_{Q,\rm eff}^{\ast} < K_{Q,\rm eff}^{\rm (cr)}$, where $K_{Q,\rm eff}^{\star} = (\sqrt{{{\beta}^{2}} + 27q_{0}^{2}K_{Q,\text{eff}}^{( \text{cr} )}( 21K_{Q,\text{eff}}^{( \text{cr} )}+32{{K}_{P}}R_h{{P}_{0}} )} - \beta) / 18$ and ${K}_{Q,\text{eff}}^{\ast} = (\sqrt{{{\beta}^{2}} + 8q_{0}^{2}K_{Q,\text{eff}}^{( \text{cr} )}( 71K_{Q,\text{eff}}^{(\text{cr})}+108{{K}_{P}}R_h{{P}_{0}} )} - \beta) / 18$ with $\beta = (16q_{0}^{2}-9)K_{Q,\text{eff}}^{( \text{cr} )}+24q_{0}^{2}{{K}_{P}}R_h{{P}_{0}}$. 
These three critical $K_{Q,\rm eff}$ lie close together, satisfying $|K_{Q,\rm eff}^{\star} - K_{Q,\rm eff}^{\rm (cr)}|/K_{Q,\rm eff}^{\rm (cr)} < 1/64$ (Fig. \ref{fig:S2S_transition}(b)). 
Numerical simulations confirm these predictions (Figs. \ref{fig:S2S_transition}(a, b); Movie S1): when $K_Q < K_Q^{\ast}$, cells remain hexagonal shape; whereas when $K_{Q} > K_{Q}^{\ast}$, cells elongate, leading to a disordered cell pattern.

An order-to-disorder transition occurs in simulations at the analytically predicted $K_Q^{\rm (cr)}$ (Eq. \eqref{eq:Critical_KQ_eff}). The hexatic order parameter, $\psi_6 = | \sum_{\alpha = 1}^{N_c} \Psi_{\alpha} / N_c|$ with $\Psi_{\alpha} = \langle \exp ({\rm i}6\theta_{\alpha\beta}) \rangle_{\beta \in \rm neighbor}$ and $\theta_{\alpha\beta} = \arg(\bm{r}_{\beta} - \bm{r}_{\alpha})$, drops as $K_Q$ exceeds $K_Q^{\rm (cr)}$ (Fig. \ref{fig:S2S_transition}(e)); such loss of order is visible in the tissue morphologies (Fig. \ref{fig:S2S_transition}(a)). 

\begin{figure}[t!]
\centering
\includegraphics[width=8.6cm]{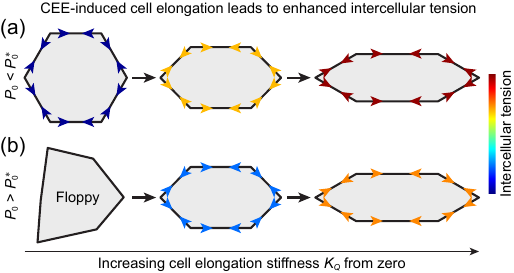}
\caption{\label{fig:mechanism} 
Schematic of the CEE‑mediated tissue solidification mechanism: (a) $P_0 < P_0^{\ast}$ case; (b) $P_0 > P_0^{\ast}$ case. 
}
\end{figure}

This order-to-disorder transition also marks a vanishing shear modulus $G_{xy}^{\rm (ST)} = {\left( {1}/{A} \right){{{\partial }^{2}}E}/{\partial \gamma _{xy}^{2}}}$, estimated at zero prestress $\gamma_{xy}=0$ (short-time limit). 
We obtain a closed-form, analytical expression for ${{G}_{xy}^{\rm (ST)}}$ (SM \cite{SM}, Sec. II) considering cell shapes under affine transformations of the regular hexagon, with principal stretches denoted $(\lambda_1,\lambda_2)$ (Fig. \ref{fig:model}(d)); $G_{xy}^{\rm (ST)}$ matches simulations well (Fig. \ref{fig:S2S_transition}(c)). 
For regular hexagonal shapes (i.e., for $K_Q < K_{Q}^{\rm (cr)}$), $\xi = \lambda_1/\lambda_2 = 1$, our analytical expression simplifies to: 
\begin{equation}
{{G}_{xy}^{\rm (ST)}} = \sqrt{3} \left({K_{Q,\rm eff}^{\rm (cr)} - {K}_{Q, \rm eff}}\right)/\left(16R_{h}^{2}\right).
\end{equation}
with $K_{Q,\rm eff}^{\rm (cr)}$ defined in Eq. (\ref{eq:Critical_KQ_eff}). 
The negative sign of the $K_{Q, \rm eff}$ term indicates a negative contribution of
CEE to the shear modulus. 
In contrast, for large cell aspect ratio $\xi = \lambda_1/\lambda_2 \gg 1$, i.e., when $K_Q  \gg K_{Q}^{\rm (cr)}$, 
\begin{equation}
G_{xy}^{\rm (ST)} \simeq \frac{16\sqrt{3}}{9} K_P \xi. \label{eq:Gxy_LargeDeformation} 
\end{equation}
Equation \eqref{eq:Gxy_LargeDeformation} demonstrates that at large cellular deformations (i.e., $\xi \gg 1$), $G_{xy}^{\rm (ST)}$ enters a linear strain-stiffening regime governed entirely by the cell perimeter stiffness $K_P$ (Fig. \ref{fig:S2S_transition}(d)). 
This stiffening stems from the increased intercellular tension $T_{\alpha\beta} = K_P (P_{\alpha} + P_{\beta} - 2 P_0)$ that CEE-mediated cell elongation induces (Fig. \ref{fig:mechanism}(a)). 
Such a strain-stiffening relation connects to work on hyperelasticity of vertex models \cite{Hernandez2023,Wang2023}. 
Overall, while $K_Q$ initially drives a negative shear modulus for a regular hexagon ($K_Q < K_{Q}^{(\rm cr)}$ regime), larger $K_Q$ ($K_{Q} > K_{Q}^{(\rm cr)}$) eventually yields a positive, finite shear modulus that restores tissue stiffness (Fig. \ref{fig:S2S_transition}(d)).

\begin{figure}[t!]
\centering
\includegraphics[width=8.0cm]{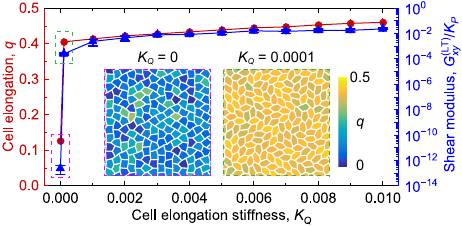}
\caption{\label{fig:FirstOrderTransition} 
First-order phase transition on increasing $K_Q$ from zero for $P_0 > P_0^{\ast}$ with $P_0 = 4$.
Average cell elongation $q$ and long-time shear modulus $G_{xy}^{\rm (LT)}$ versus cell elongation stiffness $K_Q$.
Inset: typical tissue morphologies at zero and small $K_{Q}$. 
Parameters in Table S1 \cite{SM}.
}
\end{figure}

\begin{figure}[t!]
\centering
\includegraphics[width=8.6cm]{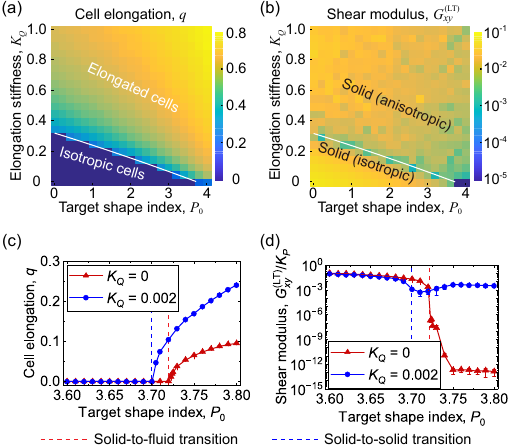}
\caption{\label{fig:PhaseDiagram} 
(a, b) Phase diagrams of (a) average cell elongation $q$ and (b) long-time shear modulus $G_{xy}^{\rm (LT)}$ in the plane of target shape index $P_0$ and cell elongation stiffness $K_Q$. 
The solid white lines refer to the theoretical prediction of the shape transition given by Eq. \eqref{eq:Critical_KQ_eff}. 
(c, d) Average cell elongation $q$ (c) and long-time shear modulus $G_{xy}^{\rm (LT)}$ (d) versus target shape index $P_0$ for two $K_Q$. 
Parameters in Table S1 \cite{SM}.  
}
\end{figure}

The long-time shear modulus $G_{xy}^{\rm (LT)}$ tracks the short-time one $G_{xy}^{\rm (ST)}$ (Fig. \ref{fig:S2S_transition}(c)): increasing $K_Q$, $G_{xy}^{\rm (LT)}$ decreases to zero at $K_Q^{\rm (cr)}$ then rise again to a plateau, see Fig. \ref{fig:PhaseDiagram}(b).
We extract $G_{xy}^{\rm (LT)}$ by applying a small simple shear strain $\gamma_{xy} = 0.01$ for a long time and relaxing to a steady state \cite{Lin2023}: $G_{xy}^{\rm (LT)} = (\langle \sigma_{xy}^{\rm tissue} \rangle_A - \langle \sigma_{xy}^{\rm tissue} \rangle_B) / \gamma_{xy}$ where $\langle \sigma_{xy}^{\rm tissue} \rangle_B$ and $\langle \sigma_{xy}^{\rm tissue} \rangle_A$ are the average tissue shear stress before and after applying the strain, respectively. 

The yield stress $\sigma_{\rm yield}$ remains finite beyond  $K_Q^{\rm (cr)}$ at which the shear modulus vanishes --- in sharp contrast with active polar \cite{Bi2016} and nematic \cite{Lin2023} CEE counterparts. We estimate $\sigma_{\rm yield}$ as the maximum shear stress along the quasistatic simple-shear curve $\sigma_{xy}(\gamma_{xy})$ \cite{Lin2023} (Movie S3). 
The yield stress remains nearly constant above $K_Q^{(\mathrm{cr})}$ and decreases only for $K_Q > K_{Q,2} \approx 0.6$ (Fig. \ref{fig:S2S_transition}(e)). This change of behavior corresponds to a transition in the yielding mode: for $K_Q < K_{Q,2}$, the stress drops abruptly at yield, indicating a brittle solid, whereas for $K_Q > K_{Q,2}$, yielding becomes smoother and ductile (Movie S3). 
This brittle-to-ductile crossover occurs in a regime where $K_Q$ also stiffens the elastic response, Fig. \ref{fig:S2S_transition}(c). Increasing $K_Q$ makes the elastic basins steeper but narrower, so that less strain is needed to reach the nearest plastic instability (Fig. \ref{fig:mechanism}(a), Movie S1). Such behavior echoes yielding phenomenology in anisotropic colloidal gels and some soft amorphous solids \cite{Shih1990,Gibaud2020,Popovic2021,Divoux2024}.

We focused so far on $P_0 < P_0^{\ast}$, but an arbitrary small $K_Q$ suffices to solidify a fluid tissue with $P_0 > P_0^{\ast}$; there, the critical is pushed to $K_Q^{\rm (cr)} = 0^{+}$, as both the cell elongation and shear moduli mark jump discontinuously (Fig. \ref{fig:FirstOrderTransition} and Movie S2). 
Considering a single cell $P_{\alpha} = P_0$ (Fig. \ref{fig:mechanism}(b)), a minute $K_Q$ leads to $P_{\alpha} > P_0$, which sets up the intercellular tension $T_{\alpha\beta} = K_P (P_{\alpha} + P_{\beta} - 2 P_0) > 0$ and solidifies the tissue \cite{Bi2015} --- thus to an immediate fluid-to-solid transition at $K_Q = 0^{+}$. Accordingly, we derive a critical $P_0$ for the solid-to-solid transition,
\begin{align}
{P}_0^{\rm (cr)} = {P}_0^{\ast} - \frac{3{K}_{Q, \rm eff}}{4{K}_P {P}_0^{\ast}} , \label{eq:Pppp}
\end{align} 
based on Eq. \eqref{eq:E_approximation} in the incompressible limit (i.e., ${K}_P \ll 1$). It shows that the CEE shifts ${P}_0^{\rm (cr)}$ toward a lower value, in agreement with simulations (SM \cite{SM}, Sec. II). 
We confirm our theoretical predictions against simulations over extensive phase diagrams in $q$ and $G_{xy}^{\rm (LT)}$ regulated by $P_0$ and $K_Q$ (Fig. \ref{fig:PhaseDiagram}). 

\textit{Robustness.} -- The CEE-mediated tissue rigidity transition is robust to (1) alternative initial cell pattern; in a random Voronoi cell pattern, we again observe a solid-to-solid transition for $P_0 < P_0^{\ast}$ and a first-order rigidity transition for $P_0 > P_0^{\ast}$ (Fig. S4), and (2) alternative values of $q_0>0$ provided $K_Q > K_{Q,\rm eff}^{\rm (cr)} / q_0^2$, as validated by simulations (Figs. S5-S8). 
For $q_0 = 0$, where cells prefer to be rounded, cell elongation elasticity restores the isotropic cell shape and tissue rigidity when $P_0 > P_0^{\ast}$ (Fig. S6), again in sharp contrast to standard vertex models.

\textit{Discussion.} --  Above the critical $K_{Q,\rm eff}$, the cell shape index exceeds $P_0^{\ast}$ (Fig.~S10), yet the long-time shear modulus $G_{xy}^{\rm (LT)}$ remains positive and finite (Fig.~S10). 
This behavior contradicts the standard vertex-model expectation that tissues fluidize once the cell shape index exceeds $P_0^{\ast}$ \cite{Staple2010,Bi2015,Park2015,Merkel2019}. 
Elongated cell shapes therefore do not necessarily imply fluid-like tissue mechanics, challenging the inference of tissue rheology from cell-shape measurements alone. This absence of fluidization also marks a sharp departure from activity-driven cell-elongation models, whether polar \cite{Bi2016,Barton2017,Lin2018} or apolar \cite{Lin2023,Sonam2023,Rozman2025}, where elongation ultimately fluidizes the tissue and triggers spontaneous flows. 

Solid-to-solid transitions -- a generic hallmark of soft condensed matter systems \cite{Du2017,Du2025} --  arise in vertex model simulations: in addition to the regular hexagonal lattice, the ground-state diagram of the Farhadifar--Staple model contains regions where square--octagon and triangle--dodecagon periodic lattices minimize the energy, with first-order transitions between these crystalline phases \cite{Farhadifar2007,Staple2010}. 
In that case, the transitions are controlled by the balance between the area elasticity and the effective line tension. Here, by contrast, the solid--solid transition is driven by an internal anisotropic mechanical stress.

\textit{Extension.} -- External cues can orient cells by coupling their shape tensors $\bm Q_{\alpha}$ to local nematic alignment field tensors $\bm M_{\alpha}$, which we capture through an effective mechanical energy, $E_{\rm align} = -\sum_{\alpha = 1}^{N_c} \chi_Q \bm M_{\alpha} : \bm Q_{\alpha}$, with $\chi_Q > 0$ quantifying the alignment strength. 
The field $\bm M_{\alpha}$ may represent mechanical loading \cite{Giverso2023}, chemical gradients \cite{Wang2023}, curvature \cite{Bell2022}, topological features \cite{Guillamat2026}, or the extracellular matrix \cite{Adar2024,Bell2025,Jacques2026}. 
With a spatially uniform $\bm{M}_{\alpha}$, increasing $\chi_Q$ drives an isotropic-to-nematic ordering of cell orientations, yielding a globally aligned tissue (Fig. S12). This transition echoes the second-order, isotropic–nematic transition of Landau–de Gennes theory for passive nematics \cite{deGennes1993}.  
The same form also captures internal, cell--cell nematic alignment, mimicking orientational elasticity in nematics \cite{deGennes1993,Marchetti2013,Giomi2015,Doostmohammadi2018,Julicher2018,Alert2020}, by taking $\bm M_{\alpha}$ as the local orientational field generated by neighboring cells, e.g., $\bm M_{\alpha} = \sum_{\beta\in\mathcal N(\alpha)} \bm Q_{\beta}$. Increasing $\chi_Q$ drives cells settling into a coherent orientation pattern (Fig. \ref{fig:model}(b)), and their spatial orientational correlations grow (Fig. S11). 
However, such local, cell-cell nematic alignment cannot overcome topological constraints to establish a global tissue orientational order. Instead, the system exhibits persistent topological defects even in the limit of large $\chi_Q$ (Fig. S11). 

\textit{Conclusion.} -- We have incorporated CEE into the vertex model, extending it to confluent tissues of autonomously elongated cells at passive mechanical equilibrium.
We show that CEE induces a solid-to-solid transition and shifts the tissue rigidity transition with the target cell shape index. 
Measuring the stiffness of strongly elongated cell cultures, such as C2C12 myoblasts or HUVECs, would directly test the link between CEE and tissue rheology.

\textit{Acknowledgments.} -- S.Z.L. acknowledges support from the National Natural Science Foundation of China (Grant No. 12502367), Fundamental Research Funds for the Central Universities, Sun Yat-sen University (Grant No. 25hytd014), Research Center for Magnetoelectric Physics of Guangdong Province (Grants 2024B0303390001), and Guangdong Provincial Key Laboratory of Magnetoelectric Physics and Devices (Grants 2022B1212010008).
J.-F.R. acknowledges support from France 2030, the French Government program managed by the French National Research Agency (ANR-16-CONV-0001) from Excellence Initiative of Aix-Marseille University - A*MIDEX.

\bibliographystyle{apsrev4-2}
\bibliography{refs}

\end{document}